\documentclass[conference]{IEEEtran}
\pdfoutput=1
\usepackage{cite}
\usepackage{amssymb}
\usepackage{algorithmic}

\usepackage{balance}
\usepackage{url}
\usepackage{color}
\usepackage{xcolor}
\usepackage{multirow}
\usepackage{caption}
\usepackage{graphicx} 

\usepackage{enumitem}
\usepackage{multicol}
\usepackage{subfloat}
\usepackage{pifont}

\usepackage{subfig,epsfig,amsfonts,amsmath,cases,amssymb,graphics, latexsym, times, algorithmic, algorithm, bbm, color, soul,wrapfig,comment}
\usepackage{epstopdf}
\usepackage{stfloats}
\usepackage{epsfig}

\def\BibTeX{{\rm B\kern-.05em{\sc i\kern-.025em b}\kern-.08em
    T\kern-.1667em\lower.7ex\hbox{E}\kern-.125emX}}

\begin{document}

\title{HealthGuard: A Machine Learning-Based Security Framework for Smart Healthcare Systems\\
}

\author {\IEEEauthorblockN{ AKM Iqtidar Newaz\textsuperscript{\textdagger}, Amit Kumar Sikder\textsuperscript{\textdagger}, Mohammad Ashiqur Rahman\textsuperscript{$\dagger\dagger$}, and A. Selcuk Uluagac\textsuperscript{\textdagger}}\\
\textit{\textsuperscript{\textdagger}Cyber-Physical Systems Security Lab, \textsuperscript{$\dagger\dagger$}Analytics for Cyber Defense Lab} \\
Department of Electrical and Computer Engineering\\
Florida International University, Miami, USA\\
\{anewa001, asikd003, marahman, suluagac\}@fiu.edu}

\maketitle

\begin{abstract}

The integration of Internet-of-Things and pervasive computing in medical devices have made the modern healthcare system "smart." Today, the function of the healthcare system is not limited to treat the patients only. With the help of implantable medical devices and wearables, Smart Healthcare System (SHS) can continuously monitor different vital signs of a patient and automatically detect and prevent critical medical conditions. However, these increasing functionalities of SHS raise several security concerns and attackers can exploit the SHS in numerous ways: they can impede normal function of the SHS, inject false data to change vital signs, and tamper a medical device to change the outcome of a medical emergency. In this paper, we propose HealthGuard, a novel machine learning-based security framework to detect malicious activities in a SHS. HealthGuard observes the vital signs of different connected devices of a SHS and correlates the vitals to understand the changes in body functions of the patient to distinguish benign and malicious activities. HealthGuard utilizes four different machine learning-based detection techniques (Artificial Neural Network, Decision Tree, Random Forest, k-Nearest Neighbor) to detect malicious activities in a SHS. We trained HealthGuard with data collected for eight different smart medical devices for twelve benign events including seven normal user activities and five disease-affected events. Furthermore, we evaluated the performance of HealthGuard against three different malicious threats. Our extensive evaluation shows that HealthGuard is an effective security framework for SHS with an accuracy of 91\% and an $F_1$ score of 90\%.

\end{abstract}

\begin{IEEEkeywords}
Healthcare, Smart Healthcare System, Anomaly Detection, Smart Medical Devices, Implantable Medical Devices, Security
\end{IEEEkeywords}

\section{Introduction}\label{sec:intro}

A rapidly growing aging population and a massive rise in the cost of healthcare have escalated the dire need of an efficient healthcare system. Indeed, according to recent figures, the global healthcare market is expected to reach \$ 53.65 billion by 2025~\cite{healthcare6}. 
%
In this fast-growing landscape, recent advancements in medical technology have led the way to 
better diagnostic tools, better treatment for patients, and devices that improve the quality of life. Specifically, with the introduction of high precision medical sensors and Internet-of-Things devices and applications, healthcare systems have become both smart and ubiquitous. Nowadays, the use of Smart Healthcare Systems (SHSs) is not limited to the medical facilities only. Besides the clinical usage, SHS also includes 
implantable and wearable medical devices to gather, store, and process various types of physiological data during daily activities of a patient~\cite{yin2018smart}.  Moreover, SHS can connect with the nearby devices or cloud (e.g., smartwatch, fitness tracker, glucose monitoring devices, etc.)  and offer a proactive approach to early detection and even prevention of medical conditions \cite{sikder2019context}. SHS also enhances the clinical and administrative workflow of healthcare organizations, and help in
the massive demand for more efficient and error-free healthcare industry.

While SHS enables many benefits with improved technology,  they unfortunately also face myriads of cybersecurity threats. Moreover, cybersecurity maturity in healthcare is  still in its early stage and healthcare data tends to be richer than financial services or retail data. 
A popular example of security concern was disabling the wireless connectivity of the pacemaker of individuals 
to protect it from hacking~\cite{intro3}. In the academic community, researchers demonstrated several cyber attacks against a commercial Implantable Cardiac Defibrillator (ICD) and Implantable Medical Devices (IMD), where an attacker could disable or reprogram the therapies performed by an ICD/IMD \cite{li2011hijacking}. However, balancing security, privacy, and usability is a challenge in the healthcare domain. Any issue concerning trustworthiness should be addressed aggressively and proactively because of the catastrophic health consequences. Thus, this area requires the immediate attention of information security research communities, medical device manufacturer and regulatory bodies. Although there are some device-specific solutions provided by the researchers, no comprehensive and centralized solution has been proposed to protect SHS from malicious threats \cite{greaves2017access}.

To address these emerging threats and shortcomings of the SHS, in this paper, we present a novel security framework, HealthGuard, to detect malicious activities in a smart healthcare system (SHS). Our framework is built upon the observation that for any change in the physiological functions of a patient, the vital signs of a specific set of medical devices changes. HealthGuard observes each device in a SHS separately and correlates the vital signs from different devices to understand the overall status of the patient. HealthGuard also uses the correlation of different body functions to differentiate normal user activities and disease-affected activities. In HealthGuard, the framework utilizes different Machine Learning (ML)-based detection algorithm (Artificial Neural Network, Decision Tree, Random Forest, k-Nearest Neighbor) to detect malicious activities in the SHS. We collected data for eight different smart medical devices and twelve benign activities (7 normal user activities and 5 disease-affected scenarios) to train HealthGuard. We also evaluated HealthGuard against three 
different threats. Our evaluation shows that HealthGuard can detect malicious activities in SHS with an accuracy of 91\% and F1-score of 90\%.
 

\vspace{6pt}
\noindent\textit{Contributions:} To summarize, our contributions are three-fold:

\begin{itemize}[nosep, wide=0pt, leftmargin=*]
    \item We present HealthGuard, a ML-based, data-driven security framework to detect threats in SHS. HealthGuard can capture the correlation between different body functions of the patient and observe the vital signs of different smart medical devices to detect malicious activities in a SHS.   
    \item We trained HealthGuard with data collected from nine databases for eight different devices and twelve benign activities including seven normal user activities and five disease-affected cases.
    \item We evaluated HealthGuard against three different threats. Our extensive evaluation illustrates that HealthGuard can detect different threats to smart healthcare system with high accuracy and F1-score.
\end{itemize}

\noindent\textit{Organization:} The rest of the paper is organized as follows: We provide an overview of security vulnerabilities in healthcare systems and existing solutions in Section~\ref{sec:related_work}. The detailed overview of the smart healthcare system is provided in Section \ref{background}. In Section~\ref{scope}, we discuss the problem scope of the current solution while focusing on this work and our considered threat models. In Section~\ref{sec:system_overview}, the detailed overview of HealthGuard is provided. We illustrate the efficiency of HealthGuard in detecting several malicious activities by analyzing several performance metrics in Section~\ref{sec:performance_evaluation}. Finally, we conclude the paper in Section~\ref{sec:conclusion}.

\section{Related Work}
  \label{sec:related_work}
 In this section, we discuss different threats to SHSs and explain the shortcomings of existing security solutions available on different platforms.

\subsection{Security Vulnerabilities}

SHS provides real-time monitoring and treatment to check the patient's health status but the functional complexity is also increasing day by day, and it makes the reliability more challenging than ever. Different works have outlined several security threats and these security threats to SHS can be categorized into four topics: hardware, software, side-channel and radio attacks. Hardware Trojan attacks have emerged as a major security concern for integrated circuits (ICs) as most ICs are manufactured in outsourced fabrication facilities. Besides for Hardware Trojan attack, there is EM radiation which can be exploited to recover internal information from the medical devices. 
EMI injection is pushed into an implantable cardiac defibrillator's sensing leads to stop it from delivering the pacing signal\cite{kune2013ghost}. Additionally, it has been shown that electromagnetic interference might be the reason for temporary or permanent malfunction in pacemakers and ICDs \cite{jilek2010safety}. 
In 2016 a ransomware attack on the Hollywood Presbyterian Medical Center shut down its network for ten days, preventing staff from accessing medical records or using medical equipment until the hospital paid the ransom about \$17,000 \cite{mansfield2016ransomware}. Security researcher of MedSec studied St. Jude Medical Maerline’s Cardiac Implantable Electronic Devices (CIEDs) and reported a "crash attack" which described a loss of radio connectivity with the CIED because of sending undisclosed radio traffic causing the CIED to stop working \cite{ransford2017cybersecurity}. 

\subsection{Existing solutions}
A high-precision, low-overhead embedded test structure (ETS) called REBEL was proposed to detect delay anomalies in Hardware Trojans (HTs) \cite{lamech2012trojan}. 
To prevent unauthorized access, limiting the communication range is an intuitive way to prevent radio attacks. 
In that scenario, near-field communication (NFC) and RFID-based channel can be utilized since they are designed for short-range communication. 
A better approach is a near-field communication (NFID) that is an extension of RFID which is gaining popularity due to its integration on mobile phones \cite{freudenthal2007suitability}.
Alkeem et al. proposed a security framework depending on sensors located on the wearable devices \cite{al2015security}. Sangpetch et al. proposed a security context framework to design the system and to evaluate security in existing systems of interest \cite{sangpetch2016security}. Abie et al. proposed a risk-based adaptive security framework using game theory and context-awareness techniques \cite{abie2012risk}. 

\subsection{Difference existing solutions}

Our framework implements an entirely new approach to detect malicious activities in SHS. There is no direct comparable solution to our work. The differences between existing solutions and HealthGuard can be noted as follows: (1) While existing solution focuses on the sensors located on the wearable devices \cite{al2015security}, HealthGuard detects malicious behaviors by considering interconnected body function. (2) HealthGuard builds a machine learning-based data-driven security framework where user identification unit for the medical devices is not required \cite{sangpetch2016security}. (3) HealthGuard does not include any overhead cost of processing complexity on the sensor node to collect data from different devices \cite{abie2012risk}. (4) Unlike threat-specific existing solutions \cite{lamech2012trojan, al2015security}, HealthGuard can detect three different types of threat in a SHS which makes it a more robust solution.

\section{Background}\label{background}

In this section, we briefly describe a SHS and different design assumptions and  features that we have considered in solution.

\subsection{Connected Smart Healthcare System}

\begin{figure}[h!]
        \centering
        \includegraphics[width=0.50\textwidth]{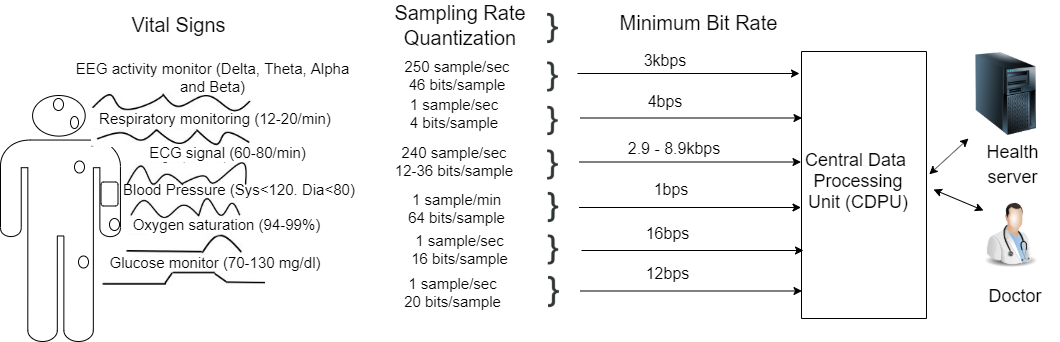}
        \caption{An example of a smart healthcare system.}
        \label{fig:background}
        \vspace{-0.5cm}
\end{figure}

SHS refers to a medical device or a group of medical devices which are equipped with different physical sensors to collect data from a patient's body and surroundings and take autonomous decisions to provide improved treatments. Also the inclusion of information and data sharing between patients and health service providers using wired or wireless technology (Bluetooth, ZigBee, etc.) improve access to care, quality of care, and increase the overall efficiency of the healthcare system. As mentioned earlier, the smart healthcare system may consist of a single or multiple smart medical devices (e.g., wearables, wireless devices, implantable devices, etc.). Here, for this work, we consider multi-device SHS. In addition to these devices, SHS also considers different non-medical parameters (e.g., the location of the patient, physical status, etc.). to provide a correct diagnosis of an event (disease, body condition, etc.). Here, different vital signs of the patient are collected by the devices to understand the overall condition of the patient. These vital signs are represented as analog signals along with their nominal values in Figure~\ref{fig:background} where vital signs are obtained, sampled and digitized for transmission as network packets to a Central Data Processing Unit (CDPU)~\cite{varshney2007pervasive}. The CDPU observes the overall condition of the patients based on the vital signs forwarded by the smart medical devices and alerts the physician in case of an emergency situation. In some instances, CDPU also takes autonomous decisions such as suggest new medicine, change the dose of the medicine, etc. For example, in Figure~\ref{fig:background}, the EEG and ECG signal monitor observe the neural activities of the patient and the heart condition, respectively. If any body function or state of the patient change, it will change ECG and EEG signal patterns. A healthcare professional can observe the change in the ECG and EEG signal and interpret this change as a cardiovascular problem. Additionally, SHS can be programmed to determine different pre-defined states (e.g., atrial problem, myocardial infraction, etc.) and provide autonomous treatment to the patient. 
\vspace{-0.1cm}
\subsection{Interconnected Body Function}

The functions of different organs in a human body are interconnected, and if any organ behaves abnormally, it affects the overall condition of the human body \cite{gray1878anatomy}. For example, if the heart rate suddenly increases in the body, it may end up causing heart palpitations, shortness of breath, etc. causing abrupt functions in other body organs. As SHS can observe different body functions at once, this interconnection can be detected and used as a feature to identify the cause of the problem. We consider this interconnection of body function as a feature of detecting anomalous behavior in SHS.  
For example, if we find a patient has high blood pressure, then the patient has cardiovascular risk factors where major risk factors are hypertension, cigarette smoking, physical inactivity, etc. Targeted organ damage are the heart, brain, chronic kidney disease, etc. Identifiable causes of hypertension are sleep apnea, drug-related problem, chronic kidney disease, etc.~\cite{chobanian2003seventh}. So in this scenario, high blood pressure of a patient can be confirmed by observing the vital signs collected from ECG, blood pressure, glucose monitoring, oxygen monitoring, sweating, EEG, and sleep monitoring device. 


\subsection{Analysis of Anomalous Behavior}

Anomalous behavior analysis refers to a model which defines all the normal behavior in a system to differentiate abnormal behavior. The capability of detecting unknown attacks makes the anomaly behavior analysis suitable for smart healthcare security framework. However, the major challenge to implementing such an analysis method in a SHS is to establish the ground truth from normal behavior with a low false positive rate. To overcome this problem, we propose an anomaly behavior analysis of the devices based on patients health condition and activities. For instance, if a person is working out in the gym, his heart-rate will rise, the oxygen level will decrease, breathing will increase, and certain brain waves will change in a pattern. To detect anomalous behavior in a SHS, a security framework should understand the ongoing activity in the human body and determine whether the activity is benign or malicious based on the vital signs collected from connected smart health devices. We consider different day-to-day user activities as well as vital signs of the body for specific diseases to understand the benign activities and identify malicious states in a SHS. For example, a person takes high cholesterol food who does not have high blood pressure suddenly gets an alarm in a blood pressure monitoring device because it crosses the minimum threshold for systolic$\ge120$. We consider both regular and disease affected scenario of the devices by observing user activities and usage patterns to build the ground truth of HealthGuard. 

\section{Problem Scope} \label{scope}
In this section, we describe the problem scope of HealthGuard using a use case scenario. Moreover, we explain different threats considered in HealthGuard that can lead to malicious activities in SHSs.

\subsection{Problem Scope}

To understand the problem scope of our work, we assume a patient (P) admits to a hospital having chest pain for the past several weeks. For emergency monitoring, a SHS is being setup with several smart medical devices observing different vital signs of the patient. For example, a cardiac monitoring ECG device, 
a pulse oximeter and an EEG is placed on P to monitor the level of oxygen on blood and neurological activity on brain respectively. We also assume that all the device is working perfectly and there is no compromised device installed in the system. Finally, the system is programmed to alert the physician and push medicine to the patient if any sudden change in the heart of the patient. At some point, the ECG started to alert the physician for the sudden drop of heart rate. However, the vital signs from EEG and pulse oximeter are normal, and the patient shows no sign of the change of heart rate. 

We introduce HealthGuard as a novel security framework that can asses the overall status of the SHS and determines whether a malicious event has occurred in the system. By using HealthGuard, several security-related questions can be answered in SHS: (1) Is the alert from one smart medical device benign or malicious? (2) The alert generated by the device is due to the influence of a disease or not? (3) Is there any outside influence (human-made or device malfunction) on the vital signs of the patient? (4) Whether an automatic decision (e.g., pushing a new dose of medicine) taken by the system is benign or not? Our proposed framework observes each smart medical device and determines the overall status of the system by connecting different vital signs of the patient. Instead of making a decision based on one device, our framework verifies the status of the patient by checking other related vitals signs. Furthermore, HealthGuard can identify a decision made by the system under the outside influence and alert the physician to prevent any erroneous treatment.

\subsection{Threat model}


HealthGuard considers malicious device behavior (e.g., an unauthorized user changing the device states) that may result in abnormal functionality of the SHS. We describe the possible attack scenarios for our work~\cite{teixeira2012attack}. It considers the attackers' system model knowledge, its disruption resources, and disclosure resources. Disruption resources enable an attacker to affect system operation and availability while disclosure resources enable the attacker to obtain sensitive information about the system during the attack by violating the data confidentiality. System model knowledge often provides an attacker with intimate details on the system to perform more complex attack.
We chose three different types of attacks that cover all three properties of attack space.

Our threat model includes (1) tampered medical devices, (2) Denial of Service (DoS) attack, and (3) false data injection. If an attacker has prior system knowledge, his disclosure resources can make false data injection attack. DoS attack makes disruption of the resource while tampered device attack can cause disruption and disclosure resources both. To better capture the threat model, we classify the threats in the following three categories:
\begin{itemize}[nosep, wide=0pt, leftmargin=*]
\item Threat 1 - Malicious Behavior 1. A malicious attacker can exist in the environment and inject forged data to perform malicious activities to change the physical condition of a patient. This threat represents false data injection in a medical device \cite{radcliffe2011hacking}.
\item Threat 2 - Malicious Behavior 2. A malicious app can be installed in any medical devices so that the device will never go into the sleep mode. This threat represents a tampered device attack \cite{williams2015cybersecurity}.
\item Threat 3 - Malicious Behavior 3. An attacker can physically be present in the environment to tamper any of the medical devices to make it temporally unavailable. This threat represents a Denial of Service (DoS) attack \cite{halperin2008pacemakers}.
\end{itemize}

Note that, any passive attack such as eavesdropping or information leakage via packet capture is considered out of the scope of HealthGuard. We also assume that the data collected from the SHS are not compromised. 

\section{System Overview}
  \label{sec:system_overview}

\begin{figure}[h!]
\vspace{-0.2cm}
        \centering
        \includegraphics[width=0.9\linewidth]{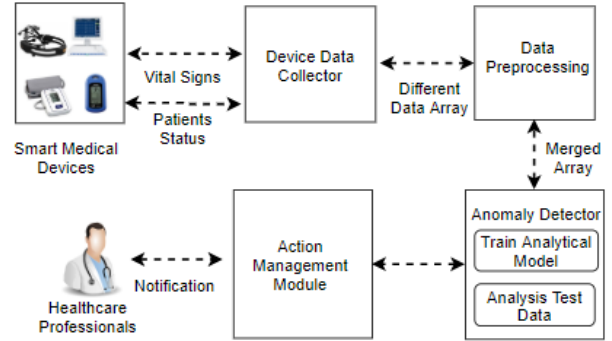}
        \caption{HealthGuard framework.}
        \label{fig:system}
\end{figure}

In this section, we present a detailed overview of HealthGuard. HealthGuard consists of four main modules: (1) data collector, (2) data preprocessing, (3) anomaly detector and (4) action management (illustrated in figure~\ref{fig:system}). The data collector module collects data from different smart medical devices. Here, each device provides a specific vital sign of the patient. These data are merged into an array in the data preprocessing module which represents the overall status of a patient at a specific time. The array generated in the data preprocessing module is fed into the anomaly detector module which decides whether or not any malicious activity is running in the SHS. Finally, the action management module notifies healthcare professionals whenever a malicious activity is detected in the SHS. The following sub-sections give details about these modules.

\subsection{Data Collector Module}

Data collector module collects data from different medical devices of a SHS. In a SHS, there can be multiple devices connected to a network and operating in a co-dependent manner. Data collector module collects the vital signs and status of the patients from these devices and saves it into a database. Based on the data collected from different devices, the collected data from each device can be represented by the following equation:

\begin{equation}
\small
Device\ Data,\ E = \{D_1, D_2, D_3, ... , D_n\}\\
\end{equation}

where $D_1$ is the set of features extracted from the device at time $t_1$, $D_2$ is the set of features extracted from the device at time $t_2$ and so on. Each device data is represented by specific data array and these data are forwarded to data preprocessing module for sampling and data merging.


\begin{table*}[]
\centering
\fontsize{12}{14}\selectfont
\resizebox{1\textwidth}{!}{
\begin{tabular}{|l|l|l|l|l|}
\hline
\multicolumn{1}{|c|}{\textbf{Device Monitoring Type}} & \multicolumn{1}{c|}{\textbf{Model}}       & \multicolumn{1}{c|}{\textbf{Feature Parameter Value}}                                & \multicolumn{1}{c|}{\textbf{Database}}            & \multicolumn{1}{c|}{\textbf{Ref.}} \\ \hline
Heart Rate and Blood Pressure                                            & QuadioArm                                 & 60-100 beats per minute, Systolic (120 mm Hg) and Diastolic (80 mm Hg)                                                                & Fetal ECG Synthetic Database, Data.Gov                      & \cite{diabetes}                                  \\ \hline
Blood Glucose                                         & MiniMed™ 670G Insulin Pump System         & 70-130 mg/dl                                                                         & UCI machine learning database of diabetes         & \cite{martin2012higher}                                  \\ \hline
Blood Oxygen                                          & iHealth Air Wireless Pulse Oximeter       & Oxygen Saturation level $\ge$ 94\%                                                   & Pattern Analysis of Oxygen Saturation Variability & \cite{bhogal2017pattern}                                  \\ \hline
Respiratory and Sweating Rate                                      & QuardioCore                               & 12-20 Breaths per minute, $0.5\mu$/min/$cm^2$                                                             & BIDMC PPG and Respiration Dataset                 & \cite{pimentel2017toward}                                  \\ \hline
Blood Alcohol                                         & Scram Continuous Alcohol Monitoring (Cam) & 0.08 g/dl                                                                            & StatCrunch dataset                                & \cite{fell2014effectiveness}                                  \\ \hline
Blood Hemoglobin                                      & Germaine AimStrip Hb Hemoglobin Meter     & 12.3-17.5 g/dl                                                                       & Hemoglobin Data in DHS Survey                     & \cite{shah2011threshold}                                  \\ \hline
Neural Activity                                       & Emotiv Insight                            & Delta (0.5-4 Hz), Theta (4-8 Hz), Alpha (8-12 Hz), Beta (16-24 Hz) \& EEG / ERP data & EEG / ERP data                                    & \cite{delorme2002single}                                  \\ \hline
Sleep and Human Motion                                                 & Fitbit Versa Smart Watch                  & REM and NREM sleep cycle                                                             & The CAP Sleep Database                            & \cite{terzano2002atlas}                                  \\ \hline
\end{tabular}}
    \caption{Devices and parameters considered for monitoring health condition. }
    \label{table:devices}
    \vspace{-0.5cm}
\end{table*}

\subsection{Data Preprocessing}

The data collector module forwards the collected data to the data preprocessing module to build the dataset of different features and aggregate them to a single array. Data collected from different devices in the data collector module consists of different patient's vital signs and status. Data merging process takes these vital signs and status to feature consideration. As different medical device has a different sampling rate, data preprocessing module sample the data according to the corresponding sampling frequency. For example, a heartbeat monitoring device measures the heartbeat of a patient in a minute (beat per minute). On the contrary, ECG monitoring device monitors the cardio-vascular state of a patient in every 10 seconds. These collected data from different devices are sampled and represented as per minute data, and data preprocessing module merges these sampled data in a single array. The data array represents the overall state of a SHS at a given time. This array can be represented as follow:

\begingroup
\small
\begin{equation}
Data\ Array,\ S = \{S_1, S_2, S_3, ... , S_n\}\\
\end{equation}
\endgroup

where $S_1, S_2, ... ..., S_n$ is the set of features extracted from $Device_1, Device_2, ... ..., Device_n$ in every minute respectively. Data preprocessing module forwards the data array to anomaly detector module for training the analytical model and detecting malicious state in the SHS.

\subsection{Anomaly Detector Module} 

Anomaly detector module uses the data arrays generated in the previous module to train different Machine Learning (ML) algorithms and detect malicious activities in the SHS. We consider two features (low computation/detection time and easy implementation) while choosing ML algorithms for HeatlthGuard. As the delay in anomaly detection can cause critical consequences to patients, low computation/detection time is a must. Again, smart healthcare devices have very low computation capability, which requires easily implementable ML algorithms in the anomaly detector. Based on these needs, we have selected the Artificial Neural Network (ANN), Decision Tree (DT), Random Forest (RF), and K-Nearest Neighbors (KNN) algorithm as these offers fast computation and easy implementation feature~\cite{sikder20176thsense, sikder2019context2}.
We briefly discuss these ML algorithms and our rationale to choose them below:
\begin{itemize}[nosep, wide=0pt, leftmargin=*]
    \item \textit{Artificial Neural Network (ANN):}  The artificial neural network is a computational model based on the functions of biological neural networks being adapted by researchers for anomaly detection. Here, the relationship among the attributes of a dataset is compared with the biological neurons, and a relationship map is created to observe the changes for each attribute \cite{linda2009neural}. We chose the Multi-layer Perceptron algorithm (MLP) for training the HealthGuard framework because our classification is multiclass, not binary classification and it is a supervised learning problem.
    
    \item \textit{Decision Tree (DT):} A decision tree uses a non-parametric modeling technique for regression and classification problems. 
    It uses divide and conquers approach and recursively select the attribute that is used to partition the training dataset into subsets until each leaf node in the tree has a uniform class membership~\cite{witten2016data}. For HealthGuard, we tested our dataset on decision tree because of the hierarchical patterns in our data set.
    
    \item \textit{Random Forest (RF):} Random forest is an ensemble classifier that uses many decision trees to model. 
    Here, a different subset of training data is selected with replacement to train each tree. We chose random forest to get more accurate and stable prediction for multi-class classification problem \cite{prinzie2008random}.
    
    \item \textit{K-Nearest Neighbor (KNN):} The K-Nearest Neighbors algorithm is instance-based learning that only stores the training samples. But does not generate a specific classification model. During classification, distances between test and training samples are calculated, and the test sample is assigned the same class label as its nearest neighbor. We chose K-nearest neighbor because it required very little training time for multi-class data sets~\cite{guney2012multiclass}.
\end{itemize}


\subsection{Action Management Module} 

The final module of HealthGuard is the action management module which notifies the healthcare professional in the event of any malicious activity in the SHS. Additionally, HealthGuard can also detect any autonomous decision taken by the system as a result of malicious activities and prevent erroneous actions of the system from avoiding any fatal consequence.


\section{Performance Evaluation}\label{sec:performance_evaluation}

In this section, we evaluate the effectiveness and feasibility of HealthGuard in detecting malicious activities in a SHS. We consider several research questions to evaluate HealthGuard in detecting malicious attacks.

\begin{itemize}
    \item[\textbf{RQ1}] What is the performance of HealthGuard in differentiating an incident occurring on disease-affected user and normal user? (Sec~\ref{normal})
    \item[\textbf{RQ2}] What is the performance of HealthGuard in detecting different malicious attacks in SHSs? (Sec~\ref{attack})
    \item[\textbf{RQ3}] What is the impact of the number of devices in SHS on the performance of HealthGuard? (Sec.~\ref{device})
    \item[\textbf{RQ4}] What is the impact of the number of attacks occurring in the SHS on the performance of the HealthGuard? (Sec.~\ref{noattack}) 
\end{itemize}
  
 \begin{table}[htbp]
\centering
\fontsize{12}{14}\selectfont
\resizebox{\columnwidth}{!}{
\begin{tabular}{|l|lllllllllll|l|}
\hline
\textbf{Disease Type}            & \textbf{ECG} & \textbf{SW} & \textbf{BP} & \textbf{GL} & \textbf{BR} & \textbf{OX} & \textbf{SL} & \textbf{HG} & \textbf{AL} & \textbf{NA} & \textbf{HM} & \textbf{Ref.} \\ \hline
High blood Pressure     & -   & \checkmark  & \checkmark  & \checkmark  & -  & \checkmark  & \checkmark  & \checkmark  & \checkmark  & \checkmark  & -  & \cite{94d682fe28184e60a9822abdddf0a2f8}    \\
High Cholesterol        & -   & \checkmark  & \checkmark  & \checkmark  & -  & \checkmark  & -  & \checkmark  & -  & \checkmark  & -  & \cite{urala2003reasons}    \\
Excessive sweating      & \checkmark   & \checkmark  & \checkmark  & \checkmark  & -  & \checkmark  & -  & \checkmark  & -  & \checkmark  & \checkmark  & \cite{heckmann2001botulinum}    \\
Abnormal oxygen level   & \checkmark   & -  & \checkmark  & \checkmark  & \checkmark  & \checkmark  & \checkmark  & -  & -  & \checkmark  & \checkmark  & \cite{bakker1991blood}    \\
High or low blood sugar & \checkmark   & \checkmark  & \checkmark  & \checkmark  & -  & \checkmark  & -  & \checkmark  & -  & \checkmark  & -  & \cite{bordia1998effect}   \\\hline
\end{tabular}}
\caption{Device status in disease affected scenarios.}
    \label{table:disease}
    \vspace{-0.5cm}
\end{table}

\subsection{Training Environment and Methodology}
  
  
To test the efficacy of HealthGuard, we collected data from eight different smart medical devices available on the Internet for different normal user activities and disease-affected users. We have selected eight smart medical devices that gave us the heart rate (ECG), blood pressure (BP), blood glucose (GL), oxygen (OX) saturation, blood hemoglobin (HG), respiratory or breathing (BR) rate, blood alcohol (AL), neural activity (NA), human motion (HM) and sleep (SL) monitoring of a person. 
We considered the threshold values of different vital signs of humans (e.g., heart rate, blood pressure, etc.) as a normal state. For example, the oxygen saturation level for a healthy person is 94-99\%, and blood hemoglobin is 12.3 - 17.5 g/dl and HealthGuard considers this range as a normal range for a user. The list of devices and a list of
selected features and sources are given in Table~\ref{table:devices}. In addition to these, we considered five different disease scenarios to understand the normal operation in disease affected scenarios of SHS thoroughly. We collected data from the selected smart medical devices for high blood pressure, high cholesterol, excessive sweating (SW), abnormal oxygen level, abnormal blood sugar. For a specific disease, a group of different but specific devices gives vital signs that are beyond the range of normal threshold. HealthGuard considers these data as disease-affected data and labels as normal operation of the SHS. The list of disease and the corresponding devices are given in Table~\ref{table:disease}.


In the training environment, we also considered seven regular user activities (sleeping, walking, exercise, stress, drunk, heart-attack, and stroke situation) and observed how the vitals from different smart medical changes with the activities. For a specific activity, a group of devices exhibits changes in the vital signs. For example, when a person does exercise his heart-rate rises, glucose, oxygen, hemoglobin level decreases, sweating increases and certain brain waves exhibit in the monitoring device~\cite{klachko1972blood}. Similarly, when a person is in stress condition, his heart-rate might rise, blood pressure increases, sweating, breathing increases and a certain part of the brain is affected~\cite{sharma2012objective}. We considered these as benign user activities and labeled as normal operation of a SHS. A detailed list of normal user activities considered in HealthGuard is given in Table~\ref{table:normal}.



\begin{table}[htbp]
\centering
\fontsize{12}{14}\selectfont
\resizebox{\columnwidth}{!}{
\begin{tabular}{|l|llllllllll|l|}
\hline
\textbf{Activity Type} & \textbf{ECG}                       & \textbf{BP}                        & \textbf{GL}                        & \textbf{BR}                        & \textbf{OX}                        & \textbf{SW}                        & \textbf{HM}                        & \textbf{HG}                        & \textbf{AL}                        & \textbf{NA}                        & \textbf{Ref.} \\ \hline
Sleeping      & \checkmark & \checkmark & \checkmark & \checkmark & \checkmark & -                         & -                         & -                         & -                         & -                         & \cite{block1979sleep}    \\
Walking       & \checkmark & -                         & \checkmark & \checkmark & \checkmark & \checkmark & \checkmark & \checkmark & -                         & \checkmark & \cite{klachko1972blood}    \\
Stress        & \checkmark & \checkmark & -                         & \checkmark & -                         & \checkmark & -                         & -                         & -                         & \checkmark & \cite{sharma2012objective}    \\
Exercise      & \checkmark & \checkmark & \checkmark & \checkmark & \checkmark & \checkmark & \checkmark & -                         & -                         & \checkmark & \cite{crabbe2004brain}    \\
Drunk         & -                         & \checkmark & \checkmark & \checkmark & -                         & -                         & -                         & -                         & \checkmark &                           & \cite{dhindsa2004differential}    \\
Heart-Attack  & \checkmark & -                         & -                         & \checkmark & -                         & \checkmark & -                         & -                         & -                         & \checkmark & \cite{chen2017brain}    \\
Stroke        & \checkmark & \checkmark & -                         & -                         & -                         & -                         & \checkmark & \checkmark & -                         & \checkmark & \cite{mackay2004atlas}  \\ \hline 
\end{tabular}}
		\caption{Device affected for normal activity pattern.}
		\label{table:normal}
		\vspace{-0.5cm}
\end{table}

For making the malicious dataset, we simulated three different attack scenarios for a SHS based on the adversary model described in Section~\ref{scope}. To perform the attack for Threat 1, we considered an attacker injected false data to medical devices to perform malicious activities. For Threat 2, we considered a malicious app installed in any of the devices which disables the sleep mode in the device (tampered device attack). For threat 3, we simulated a scenario to impede the normal operation of a smart medical device which illustrates the DoS attack in the system. We built a simulation environment in MATLAB using digital signal processing toolbox and Poisson distribution to randomize the attack scenario. We used Poisson distribution to describe the attack scenarios as rare events in a large dataset. 

To test HealthGuard, we collected 20,000 data instances in total where 17,000 were for healthy and disease infected people, and 3,000 were simulated attack data in dataset. Next, we divided our data into two sections, where  70\% of the collected dataset was used to train the framework, and 30\% of the collected data along with malicious dataset were used for testing purpose \cite{amazon}.

  
\begin{table}[htbp]
\centering
\fontsize{12}{14}\selectfont
\resizebox{\columnwidth}{!}{
\begin{tabular}{|l|llll|llll|}
\hline
                   & \multicolumn{4}{c|}{\textbf{Benign}}                    & \multicolumn{4}{c|}{\textbf{Malicious}}                 \\ \cline{2-9} 
                   & \textbf{KNN} & \textbf{DT} & \textbf{RF} & \textbf{ANN} & \textbf{KNN} & \textbf{DT} & \textbf{RF} & \textbf{ANN} \\ \hline
\textbf{Accuracy}  & 0.903        & 0.931       & 0.898       & 0.927        & 0.878        & 0.909       & 0.865       & 0.910        \\
\textbf{Precision} & 0.90         & 0.92        & 0.90        & 0.92         & 0.88         & 0.91        & 0.86        & 0.90         \\
\textbf{Recall}    & 0.90         & 0.93        & 0.90        & 0.93         & 0.88         & 0.91        & 0.86        & 0.91         \\
\textbf{F1-score}  & 0.90         & 0.93        & 0.90        & 0.93         & 0.87         & 0.90        & 0.86        & 0.89         \\ \hline
\end{tabular}}
    \caption{Performance of HealthGuard in detecting benign and malicious events in SHS.}
    \label{table:benign and malicious}
    \vspace{-0.7cm}
\end{table}

\subsection{Performance Metric}
  
In the evaluation of HealthGuard, we used four different performance metrics: Accuracy, Precision, Recall, F1-score. Accuracy refers to the degree of closeness of a measured quality to that quality's true value, and Precision calculates the fraction of correct positive identifications. Recall identifies the portion of correctly identified positives. $F_1$-score measures a test's accuracy considering both precision and recall. 



\subsection{Evaluation with only disease-affected and normal activities} \label{normal}
In a SHS, different benign but rare events can occur as an effect of a disease in the patient's body and user activities (e.g., sleeping, working out, etc.). A security framework should be able to detect these diverse types of events correctly. To evaluate the performance of HealthGuard in detecting benign activities, we chose twelve different scenarios in total including vitals collected for seven user activities and five disease-affected scenarios. Table~\ref{table:benign and malicious} presents the evaluation results associated with different benign activities. We can observe that accuracy, and $F_1$score varies from 90-93\% for different algorithms. We achieved the highest accuracy and $F_1$score of 93\% using the DT algorithm. We can also observe that HealthGuard achieved the lowest accuracy of 89\% using the RF algorithm. The accuracy of KNN and ANN are 90.3\% and 92.7\% respectively. In summary, HealthGuard can achieve the highest accuracy and $F_1$score using a decision tree algorithm for correctly identifying benign activities.

\begin{table*}[t!]
\centering
\fontsize{12}{15}\selectfont
\resizebox{1\textwidth}{!}{
\begin{tabular}{|l|llll|llll|llll|llll|llll|}
\hline
\textbf{Device Count} & \multicolumn{4}{c|}{\textbf{4}}                                              & \multicolumn{4}{c|}{\textbf{5}}                                              & \multicolumn{4}{c|}{\textbf{6}}                                              & \multicolumn{4}{c|}{\textbf{7}}                                              & \multicolumn{4}{c|}{\textbf{8}}                                              \\ \hline
\textbf{Algorithm}    & \textbf{Accuracy} & \textbf{Precision} & \textbf{Recall} & \textbf{F1-score} & \textbf{Accuracy} & \textbf{Precision} & \textbf{Recall} & \textbf{F1-score} & \textbf{Accuracy} & \textbf{Precision} & \textbf{Recall} & \textbf{F1-score} & \textbf{Accuracy} & \textbf{Precision} & \textbf{Recall} & \textbf{F1-score} & \textbf{Accuracy} & \textbf{Precision} & \textbf{Recall} & \textbf{F1-score} \\ \hline
\textbf{KNN}          & 0.812             & 0.82               & 0.81            & 0.78              & 0.823             & 0.81               & 0.82            & 0.79              & 0.845             & 0.84               & 0.84            & 0.83              & 0.878             & 0.88               & 0.88            & 0.87              & 0.878             & 0.88               & 0.88            & 0.87              \\
\textbf{DT}           & 0.839             & 0.83               & 0.84            & 0.81              & 0.850             & 0.85               & 0.85            & 0.82              & 0.866             & 0.86               & 0.87            & 0.85              & 0.909             & 0.91               & 0.91            & 0.90              & 0.909             & 0.91               & 0.91            & 0.90              \\
\textbf{RF}           & 0.772             & 0.75               & 0.77            & 0.76              & 0.778             & 0.75               & 0.78            & 0.76              & 0.804             & 0.79               & 0.80            & 0.79              & 0.865             & 0.86               & 0.86            & 0.86              & 0.865             & 0.86               & 0.86            & 0.86              \\
\textbf{ANN}          & 0.811             & 0.78               & 0.81            & 0.77              & 0.828             & 0.82               & 0.83            & 0.79              & 0.861             & 0.82               & 0.86            & 0.82              & 0.9111            & 0.89               & 0.91            & 0.89              & 0.910             & 0.90               & 0.91            & 0.89              \\ \hline
\end{tabular}}
    \caption{Performance evaluation of HealthGuard considering different number of devices.}
    \label{table:three}
    \vspace{-0.5cm}
\end{table*}

\subsection{Evaluation with Different Attack Scenarios} \label{attack}
To evaluate HealthGuard against different malicious attacks, We considered three different threats (tampered device, DoS, and false data injection attack) in a SHS. We collected 3000 different instances for these attack scenarios and tested HealthGuard. 
From Table~\ref{table:benign and malicious}, we can observe the ANN algorithm performs with the highest accuracy and $F_1$score of 91\% and 89\% respectively. For the DT algorithm, one can notice that the accuracy decreases to 90\% while
$F_1$score increases slightly (90\%). For KNN and RF, accuracy and $F_1$score vary from 86-87\%. In short, the ANN algorithm performs better in detecting different malicious activities in SHS.

\begin{figure*}[t!]
\vspace{-0.1in}
  \centering
  \subfloat[]{\includegraphics[width=0.25\textwidth]{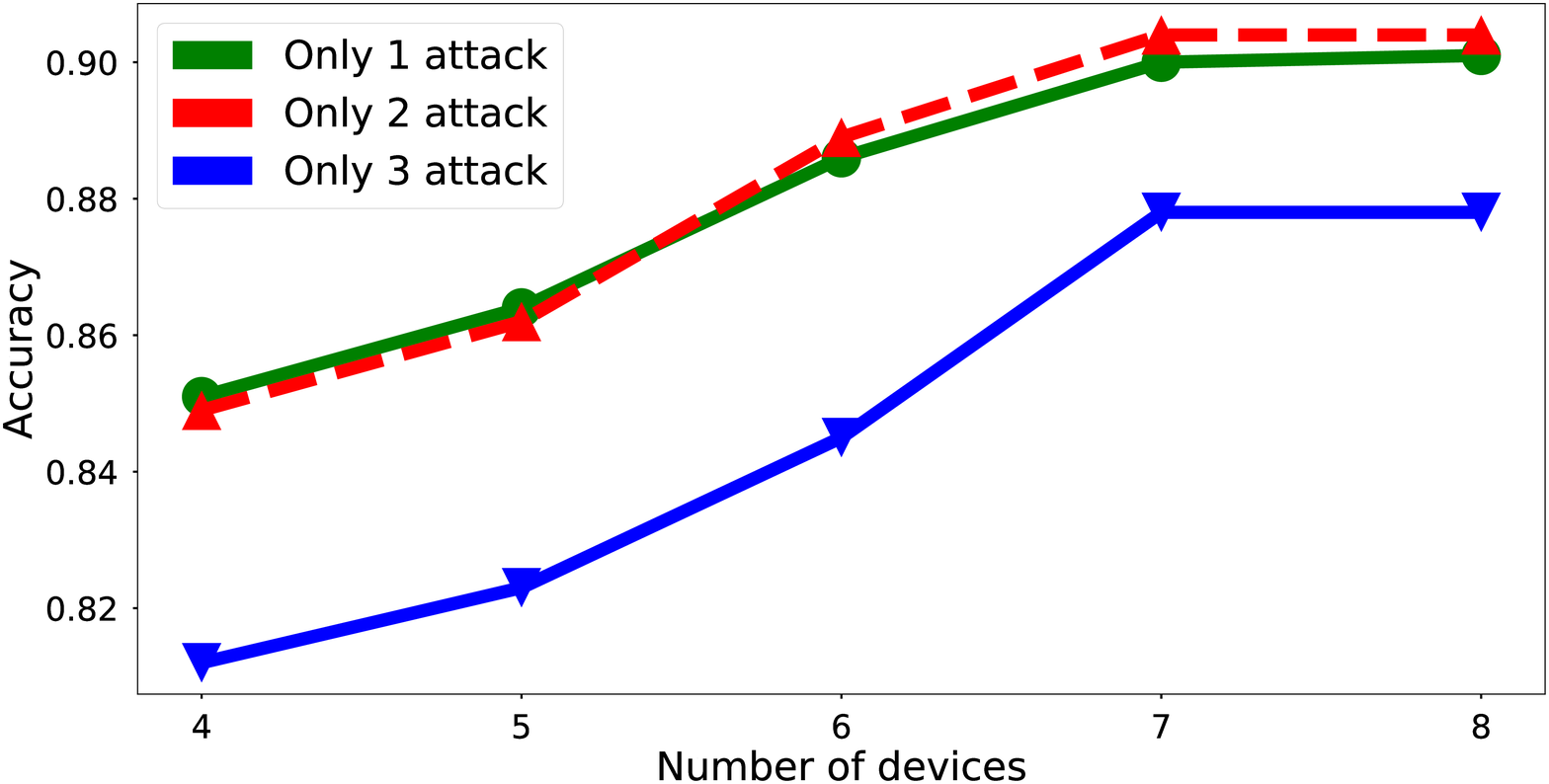}\label{fig:f8}}
\subfloat[]{\includegraphics[width=0.25\textwidth]{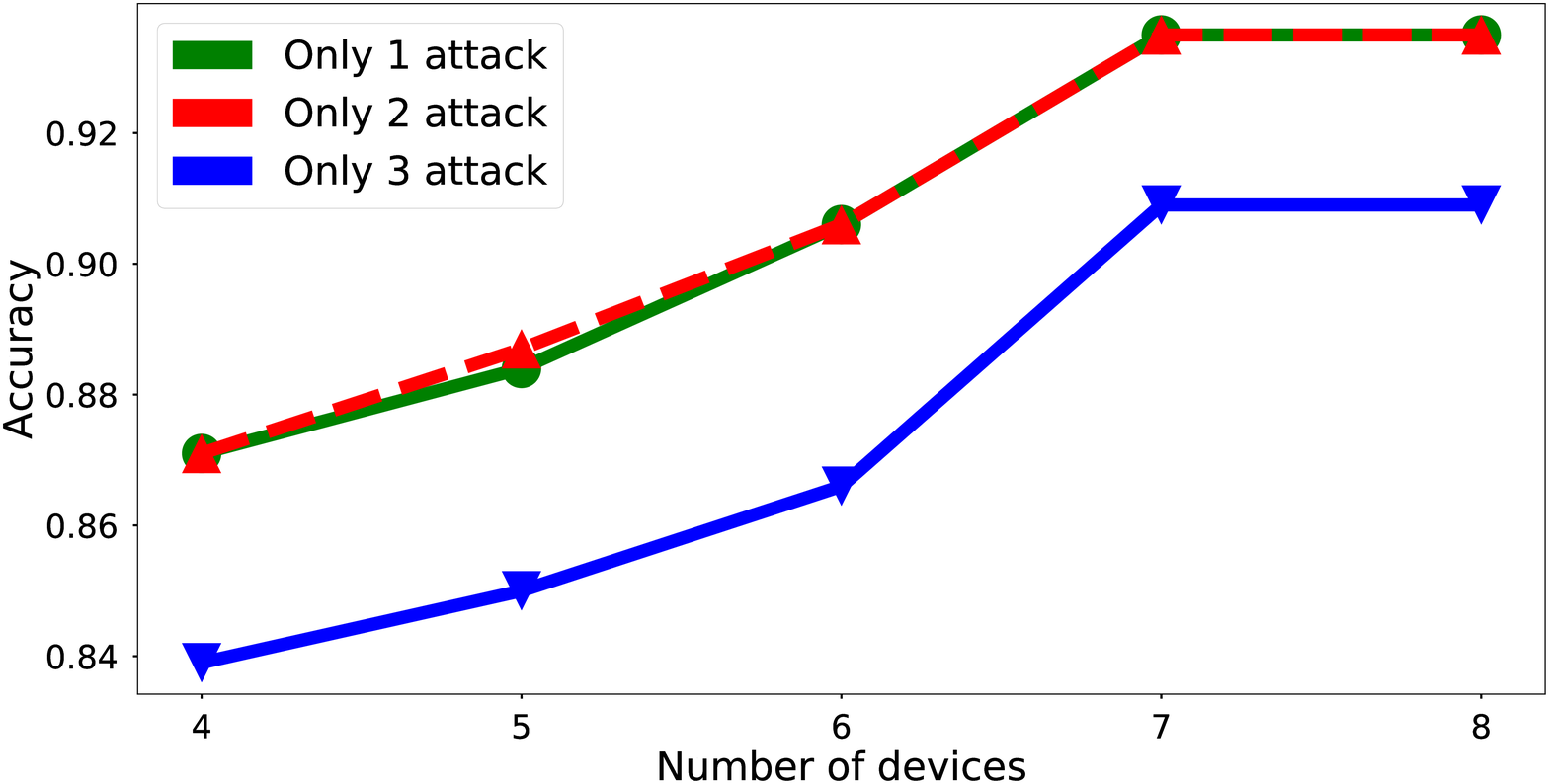}\label{fig:f9}}
\subfloat[]{\includegraphics[width=0.25\textwidth]{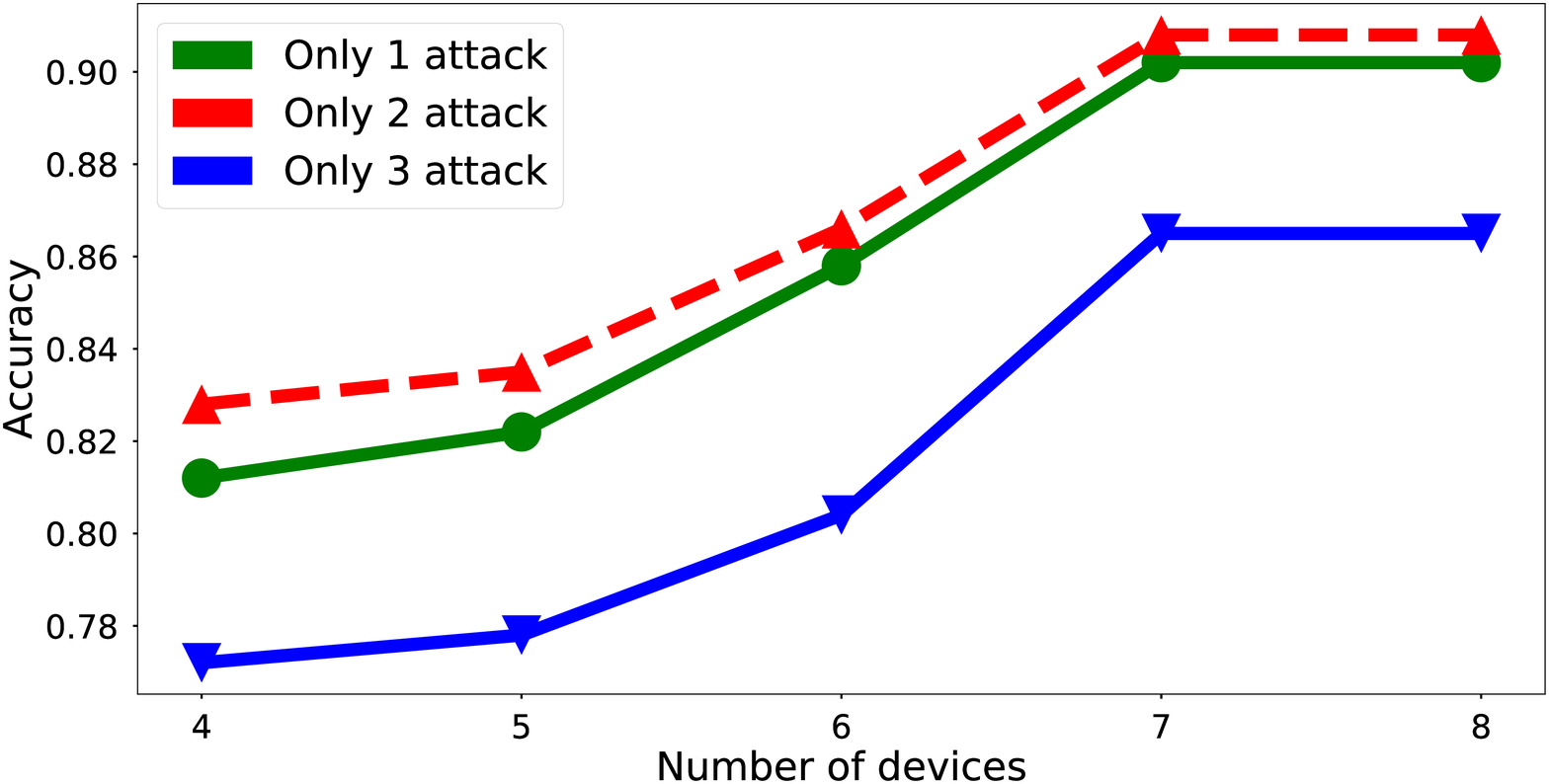}\label{FP}}
\subfloat[]{\includegraphics[width=0.25\textwidth]{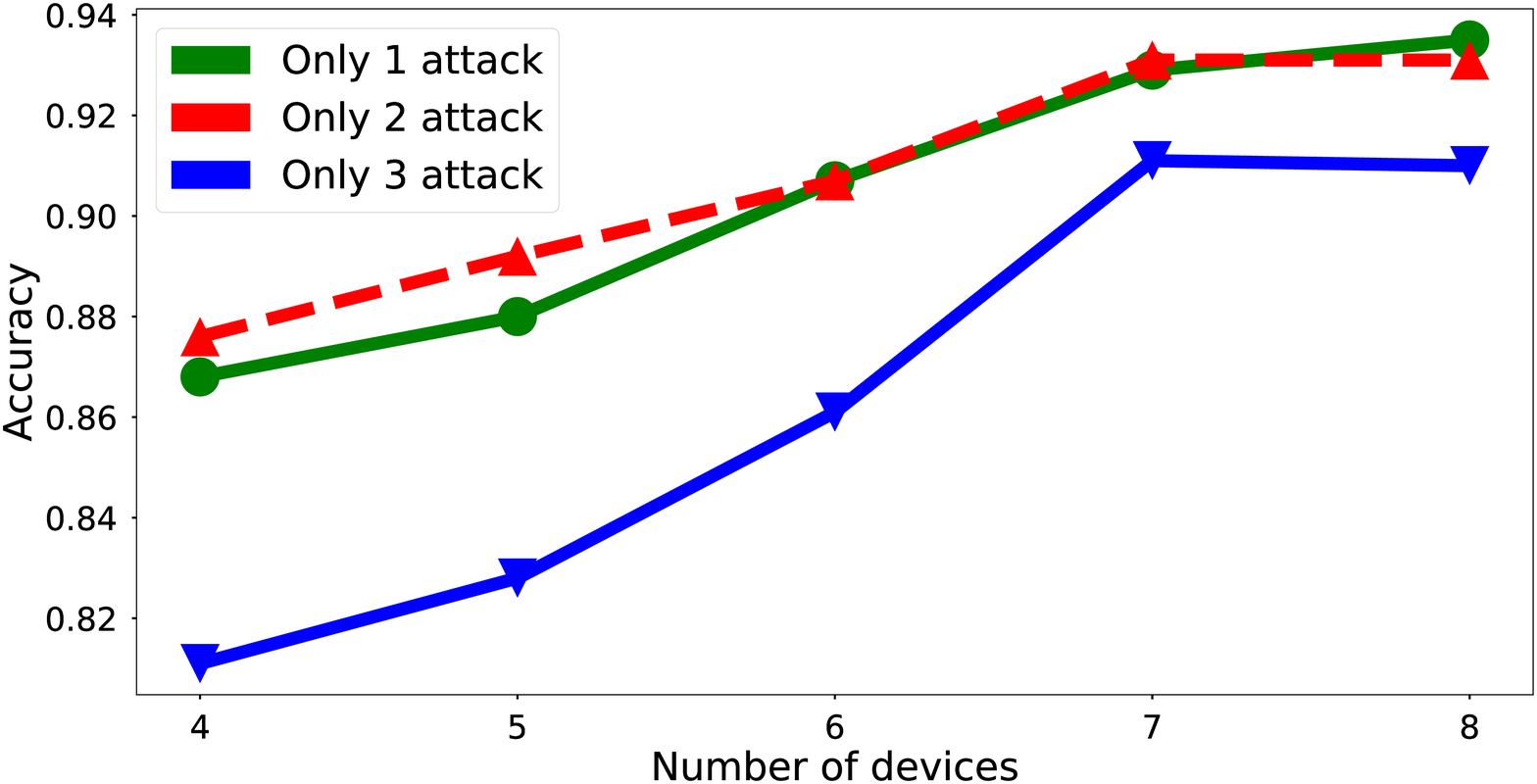}\label{acc}}
\vspace{-0.2cm}
      \caption{Accuracy of HealthGuard for different ML algorithm (a) KNN, (b) DT, (c) RF, (d) ANN.}
     \label{fig:accuracy}
      \vspace{-0.5cm}
\end{figure*}

\subsection{Evaluation with Different Number of Devices}\label{device}
In a SHS, a different number of smart medical devices can be connected to monitor the overall status of a patient or user. To test the effect of the number of connected devices in HealthGuard, we change the configuration of SHS and performed the detailed evaluation (Table~\ref{table:three}). We can observe that the accuracy and
$F_1$score decreases with the number of connected devices in SHS. HealthGuard performs better if the SHS has more devices connected in the system as it is easier to detect an event correctly with more vital signs giving information about the patient/user. We can observe that ANN performs the best with the accuracy and $F_1$score range of 81-91\% and 77-89\% for 4 and 8 connected devices respectively. We can also notice that DT is less affected for decreasing the number of devices as the accuracy and $F_1$score change only 7\% and 9\% respectively. In summary, HealthGuard achieves the highest accuracy and $F_1$score using ANN, however, the performance is less affected in DT with a decreasing number of devices in SHS.




\subsection{Evaluation with Simultaneous Attack Scenarios} \label{noattack}
A SHS can be vulnerable to multiple malicious attacks at the same time. To understand the effect of multiple simultaneous attacks on HealthGuard, we simulated different simultaneous attacks using Poisson distribution. Figure~\ref{fig:accuracy} illustrates the impact of simultaneous attacks on HealthGuard. We can notice that all the detection mechanism performs with the highest accuracy if only one attack is active in the system. The accuracy decreases with the number of attacks in the system. We can also observe that ANN performs with the highest accuracy in all three scenarios (one attack, two attacks, and three attacks). HealthGuard can achieve 93\% accuracy for one and two simultaneous attacks and 91\% accuracy for three simultaneous attacks using ANN.

\section{Conclusion}\label{sec:conclusion}

Smart Healthcare Systems offer better diagnostic tools and treatment for patients, but they also raise 
many security concerns, as discussed in this work. 
To address these security concerns, in this paper, we presented HealthGuard, a novel machine-learning-based security framework that can assess the overall status of a SHS and can determine if there is any malicious activity has occured in the system. We evaluated HealthGuard in multiple medical settings 
considering healthy and disease-infected people in a SHS. Moreover, 
HealthGuard is highly effective and efficient in detecting several threats. Specifically, HealthGuard can achieve 91\% accuracy in detecting different attacks. 

\section{Acknowledgment}\label{sec:acknowledgment}

This work was partially supported by US NSF-CAREER-CNS-1453647, and Florida Center for Cybersecurity Capacity Building Program.  The  views  expressed are  those  of  the authors only.

\bibliographystyle{IEEEtran}
\bibliography{reference.bib}

\end{document}